%% file: main.tex
\documentclass[12pt]{article}
\usepackage[utf8]{inputenc}
\usepackage{charter}
\usepackage{fullpage}
\usepackage{graphicx}
\usepackage{lipsum}
\usepackage[sort&compress,numbers]{natbib}
\usepackage[total={6.5in,9in},top=1in,headsep=0.1in,headheight=1in]{geometry}

\usepackage[hyphens]{url}

\usepackage[hyphenbreaks]{breakurl}

\usepackage[hyperindex,breaklinks]{hyperref}
\usepackage{xurl}
 \hypersetup{
           breaklinks=true,   
           colorlinks=true,   
           pdfusetitle=true,  
                    }

\usepackage{amsmath}
\usepackage{xcolor}
\usepackage{verbatim}
\usepackage{multirow}
\usepackage{orcidlink}

\definecolor{hookgreen}{rgb}{0.0,0.44,0.0}
\definecolor{periwinkle}{RGB}{104,127,243}
\definecolor{eggplant}{RGB}{83,27,147}
\definecolor{olive}{RGB}{74,80,7}

\newcommand\snowmass{
\begin{center}
  \rule[-0.2in]{\hsize}{0.01in}\\
  \rule{\hsize}{0.01in}\\
  Submitted to the Proceedings of the US Community Study\\ 
  on the Future of Particle Physics (Snowmass 2021)\\
    \vskip 0.05in
    {\it Snowmass2021 CEF03 Diversity \& Inclusion}\\
  \rule{\hsize}{0.01in}\\
  \rule[+0.2in]{\hsize}{0.01in}\\[-2em]
\end{center}
}

\usepackage[firstpage=true]{background}
\backgroundsetup{contents={\parbox{6.5in}{\snowmass}}, scale=1,placement=top,opacity=1,color=black,position={3.25in,1.2in}}

\usepackage{fancyhdr}
\fancypagestyle{plain}{%
  \fancyhf{}%
  \fancyhead[C]{}
  \fancyfoot[C]{\thepage}
}

\fancypagestyle{empty}{%
  \fancyhf{}%
  \fancyhead[C]{{\it Snowmass2021 CEF03 Diversity \& Inclusion}}
  \fancyfoot[C]{\thepage}
}
\title{Building a Culture of Equitable Access and Success for Marginalized Members in Today's Particle Physics Community}

\date{}

\usepackage{authblk}

\author[1]{K\'et\'evi A. Assamagan
\orcidlink{0000-0002-4826-2662}}
\affil[1]{Physics Department, Brookhaven National Laboratory, Upton, NY}

\author[2,3]{Olivia M. Bitter
\orcidlink{0000-0002-0600-9095}}
\affil[2]{Fermi National Accelerator Laboratory, Batavia, IL}
\affil[3]{The University of Chicago, Chicago, IL}

\author[4]{Mu-Chun Chen
\orcidlink{0000-0002-5749-2566}}
\affil[4]{Department of Physics and Astronomy, University of California, Irvine}

\author[5]{Ami Choi
\orcidlink{0000-0002-5636-233X}}
\affil[5]{California Institute of Technology}

\author[2]{Jessica Esquivel
\orcidlink{0000-0003-2398-7293}}

\author[6]{Kathryn Jepsen
\orcidlink{0000-0002-0382-5126}}
\affil[6]{SLAC National Accelerator Laboratory}

\author[7,8]{Tiffany R. Lewis \orcidlink{0000-0002-9854-1432}}
\affil[7]{Astroparticle Physics Laboratory, NASA Goddard Space Flight Center, Greenbelt, MD}
\affil[8]{NASA Postdoctoral Program Fellow}

\author[9]{Azwinndini Muronga
\orcidlink{0000-0003-3501-5272}}
\affil[9]{Faculty of Science, Nelson Mandela University, Gqeberha, South Africa}

\author[2]{Fernanda Psihas
\orcidlink{0000-0001-7393-4662}}

\author[11]{Lucianne Walkowicz
\orcidlink{0000-0003-2918-8687}}
\affil[11]{The JustSpace Alliance}

\author[12]{Yuanyuan Zhang
\orcidlink{0000-0001-5969-4631}}
\affil[12]{Department of Physics and Astronomy, Texas A\&M University, College Station, TX}

\begin{document}

\maketitle

\begin{abstract}
Diversity, Equity, Inclusion, and Accessibility (DEIA) are not only called for to ensure morality and justice in our society, 
they also support ongoing and future excellence in particle physics. Over the past decade, the particle physics community has devised programs to support DEIA along multiple axes, and the way we think about measuring and implementing these initiatives has evolved. DEIA in physics is a broad topic, so in this paper we focus on the experiences of marginalized communities 
and outline ways different stakeholders can build a culture of equitable access for the success of marginalized individuals. Specifically, we identify urgent needs in the following areas: (1) We need to acquire a better understanding of the status quo, both quantitatively and qualitatively, to assess the effectiveness of existing programs and to develop best practices; (2) we need to develop effective and inclusive ways to engage marginalized communities; (3) we need to create infrastructure to better support members of marginalized communities, on an academic, financial and personal level; (4) we need to create an environment conducive to equitable access and success by establishing community expectations, fostering inclusion in social interactions, and holding individuals and institutions accountable; and (5) we need to establish a mechanism to monitor progress in the area of DEIA, including the implementation of the recommendations enumerated in this paper and others during the Snowmass 2021 process.  

\end{abstract}

\clearpage

\tableofcontents


\setcounter{footnote}{0}


\clearpage
\sloppy{
\include{introduction}

\include{status}
\include{engage}
\include{pathway}

\include{culture}
\include{recommendations}

\section*{Acknowledgements}
We are grateful for the crucial support Daria Wang has provided throughout the entire process. We would also like to thank Ed Bertschinger for his important feedback on the draft. In addition, we acknowledge the useful conversations with many particle physics community members as well as Town Hall panelists from other disciplines, which have helped to shape the direction of the white paper. Tiffany Lewis acknowledges support from the NASA Postdoctoral Program at NASA Goddard Space Flight Center, administered by Oak Ridge Associated Universities. M.-C.C. was supported in part by National Science Foundation under Grant No. PHY-1915005. AC acknowledges support from the SPHEREx project under a contract from NASA/Goddard Space Flight Center to the California Institute of Technology.
}

\bibliographystyle{JHEP}
\bibliography{refs}
\end{document}

%% file: introduction.tex
\section{Why is Removing Marginalization Important in Particle Physics?}

Scientific excellence is achieved by people, and people are at their best when they are treated well. The only way to ensure our scientific community reaches its full potential is to ensure that \textit{every single one} of its members has equal access to resources, education, and career opportunities. It falls to organizations, institutions, and funding agencies to provide these resources and opportunities, and to foster a culture of inclusion in which people can succeed. 

Diversity, equity, inclusion and accessibility drive intellectual excellence. The particle physics community and its stakeholders should reflect on how barriers to the pursuit of a scientific career shape the population responsible for scientific progress, and how that progress may be limited by exclusion. 

\subsection{Marginalized Communities in Particle Physics}
Marginalized communities are groups of people who experience a sense of being pushed to the edge of a community or being excluded from important groups and activities. Those who are marginalized are often members of what are called underrepresented minority (URM) groups\footnote{We suggest caution when using this term, as the language literally implies ``less than" and thus may further harm the groups who are labeled this way. See, e.g., \cite{bensimon2016misbegotten,williams2020underrepresented} who argue that the term is both dehumanizing when used to refer to racial and ethnic groups and hides inequalities across different groups when used as an umbrella term. Here, we aim to be clear about who we are referring to and also use this term when describing literature or statistical studies that use it.}. But there are differences between what is meant by ``marginalized," ``underrepresented," and ``minority." For example, on average between 2014-2018, Black students made up 15.64\% of the college-age population in the United States but earned just 3\% of bachelor's degrees in physics awarded in the country~\cite{APS_AA_bachelors}. Black students were in the minority, as they made up less than 50\% of the college-age population. They were also underrepresented, as the percentage of Black students earning bachelor's degrees in physics was lower than the percentage of Black students in the overall US college-age population. Black students could thus be described as an underrepresented minority among those earning bachelor's degrees in physics in 2018. 

One does not need to be a URM to be marginalized, but URMs do often face marginalization. One reason Black students are underrepresented in physics is due to the marginalization that Black students face. 

In particle physics, as in society at large, people are marginalized based on factors such as their:
\begin{itemize}
\itemsep0em 
\item gender identity and sexual orientation; 
\item race and ethnicity; 
\item disability (both visible and invisible)~\cite{Assamagan:2022oka}; 
\item socioeconomic status; 
\item status as a first-generation college student; 
\item immigration status; 
\item veteran status. 
\end{itemize}

Dedicated discussions of the unique issues experienced by marginalized communities can be found in the referenced white papers.

An individual can face marginalization due to more than one aspect of their identity, and an individual can face marginalization based on some aspects of their identity and privilege based on others. The concept of intersectionality~\citep{crenshaw1989demarginalizing} is the recognition that multiple aspects of a person's identity overlap and interact to shape their experience of the world. For example, a Vietnamese American woman in STEM~\citep{Vietnamese_women_stem} may endure sexism that is unique to her ethnicity and racial group and related to her socioeconomic status. 

Women, people of color, foreign nationals, immigrants, and people who do not conform to mainstream societal norms have historically been excluded from education and career opportunities throughout academia, but the trend has been especially persistent in particle physics. The persistence of these conditions is the fundamental reason it is difficult to recruit and retain diverse participants in the field.

The difficulties that the under-represented and marginalized face in our field often permeate their day-to-day experience, affecting their ability to succeed significantly with respect to those who are not marginalized. These effects are often accentuated when such difficulties are either invisible or poorly understood. In the case of neurodiverse individuals, for example, the marginalization they experience will often hinder their ability to succeed both academically and socially without clear signs that they are being caused by the environment as opposed to a lack of ability or capacity.

\subsection{Diversity, Equity and Inclusion as a National Priority in HEP}

Diversifying the workforce in HEP (High-Energy Physics) has for several years been a priority for the U.S. Department of Energy's Office of Science.

The 2016 Report of the Committee of Visitors to the High Energy Physics Advisory Panel~\cite{CoV_report_2016} warned that “[u]nder-representation of women and minorities in physics as a whole continues to be a challenge. Greater attention should be paid to promoting an inclusive environment in order to provide encouragement to research groups to improve the diversity of the HEP workforce. HEP review processes for university groups and laboratories should consider activities that promote diversity and inclusion in the workforce and the workplace.”

In their subsequent report~\cite{CoV_report_2020}, in 2020, the Committee of Visitors commended HEP “for their considerable effort to address diversity, equity, and inclusion (DEI),” writing that they “express significant awareness of the issues and their importance.” The committee recommended HEP “develop and implement strategies and policies to foster diversity, equity, and inclusion in supported university groups as well as at the laboratories.” The report praised HEP for adding “mentorship and other diversity considerations” to the “Program Policy Factors” that add to the strength of a grant proposal.

DOE's Office of Science has put resources toward the goal of developing a diverse workforce in HEP.
The Office of Science’s Office of Workforce Development for Teachers and Students~\cite{WDTS_website} “sponsor[s] workforce training programs designed to motivate students and educators to pursue careers that will contribute to the Office of Science’s mission in discovery science and science for the national need.” The Office of Science website~\cite{WDTS_website_about} states that: “The mission of [WDTS] is to help ensure that DOE and the Nation have a sustained pipeline of highly skilled and diverse science, technology, engineering, and mathematics (STEM) workers.”
In addition, the Office of Science’s Office of Scientific Workforce Diversity, Equity, and Inclusion \cite{SW-DEI_website} “collaborates across [the Office of Science] to advance organizational best practices…to promote DEI at the SC DOE National Laboratories.”
In 2020, DOE Office of Science published its ``Summary of DOE Office of Science Recognized and Promising Practices for DOE National Laboratory Diversity, Equity, and Inclusion Efforts"~\cite{SC_DEI_effort}.

\subsection{Engagement with Marginalized Communities since Snowmass 2013}

The 2013 Snowmass Report~\cite{2013snowmass} mentions DEI with only minimal discussion. Terms like ``equity", ``inclusion", and ``marginalized" (or their variations) are not used in societal contexts. The terms ``diversity" and ``underrepresented minorities" make a brief appearance in the Communication, Education, and Outreach report (Section 6 of~\cite{2013snowmass} and~\cite{2014arXiv1401.6119B}), focusing on K-12 education. The text notes that ``the number of students from underrepresented minorities in the field is still too low for meaningful statistics" and recommends partnering with national efforts to engage historically underrepresented students in elementary school. However, recruitment is only one important piece of a much bigger landscape where change is needed. 

Overall, the 2013 Snowmass Report~\cite{2013snowmass} lacks clear discussion of \textit{who} comprises these ``underrepresented minorities", the \textit{circumstances} that may cause marginalization of some groups of physicists, or any \textit{recommendations} related to retention of existing minoritized and/or marginalized members of our community\footnote{The subgroup report on Communication with Teachers and Students~\cite{snowmass2013subgroupcomm} provides a table specifying fractions for women, Asian Americans, Hispanic Americans, and African Americans for high school and college physics that did not make it into the main summary report.}. Thus, Snowmass 2021 will serve as a baseline for discussions and activities specific to marginalized groups in particle physics.

The 2020 murders of Ahmaud Arbery, Breonna Taylor, George Floyd, and too many others ignited new action and awareness worldwide and within our particle physics community. Shutdown STEM\footnote{\href{https://www.shutdownstem.com/}{https://www.shutdownstem.com/}}, Particles for Justice\footnote{\href{https://www.particlesforjustice.org/}{https://www.particlesforjustice.org/}}, and Vanguard: Conversations with Women of Color in STEM\footnote{\href{https://www.vanguardstem.com/}{https://www.vanguardstem.com/}} led movements (e.g., Strike for Black Lives on June 10, 2020) that drew new attention to our urgent responsibility to fight racism and injustice within our academic and research institutions. 

Resources that outline evidence-based solutions \cite[e.g.,][]{teamup,changenow} are emerging, but these movements and resources have some poignant limitations. Historically, they have been led and developed by physicists who are themselves marginalized. And evaluation is often limited by the lack of comprehensive demographic data. 

It is important for our community to sustain and expand the momentum it developed in 2020. It is important for our community to set up a mechanism to monitor progress toward diversity, equity, inclusion and accessibility from now through the next Snowmass process, and beyond.   


\subsection{Advantages to the Community of a Commitment to Diversity}

Beyond the ethical reasons that justify the pursuit for an equitable and diverse environment, there are significant benefits that our field stands to gain from a true commitment to diversity, equity, and inclusion. Research on this subject indicates that environments where diversity is celebrated and productive dissent is encouraged often  mitigate the impact of groupthink, encourages productive criticism, and enable a culture of creativity and innovation in the work environment~\cite{FP1,FP2,FP3}. Original thinking, creativity, and innovative problem-solving are all catalysts of discovery science. Thus, the scientific endeavor stands to make significantly from a truly diverse workforce. 

One area of DEIA that can especially contribute to developing an increasingly innovative environment in HEP is our attention to neurodiversity.  It is undeniable that differences in cognition -- often understood as hidden disabilities -- present significant limitations to those afflicted by them, specifically in realms of behavior and achievement which are natural to their neurotypical counterparts. However, research into these same cognitive differences has revealed a multitude of advantages in the performance of specific tasks and especially in the areas of innovation, creativity, error assessment, intense focus, and personal resilience, among others~\cite{ADHD, Autism}. Efforts to increase neuriodiversity in the workplace are increasingly prevalent in industry, with tech companies and investment firms leading the way with targeted efforts towards hiring neurodiverse individuals~\cite{ND1,ND2}.

%% file: status.tex
\section{Understanding the Status Quo}

The March 2022 article in \textit{Science} ``A push for inclusive data collection in STEM organizations"~\cite{burnett2022push} explains that ``[p]rofessional organizations in science, technology, engineering, and mathematics (STEM) are well-positioned to improve the recruitment and retention (R\&R) of underrepresented groups by providing targeted professional development, networking opportunities, and political advocacy. Tailoring these initiatives to specific underrepresented groups can enhance their impact, but this is predicated on organizations knowing their demographic make-up." 

\subsection{Current Representation of Marginalized Groups}

To improve the representation and experiences of marginalized scientists, it can be helpful to state plainly where diversity in particle physics currently stands, along multiple axes.

\paragraph{Data on Race \& Ethnicity:}
The American Physical Society (APS) and American Institute of Physics (AIP) \cite{APS_AIP_stats} collect data from the Integrated Post-secondary Education Data System, which collects its data from the National Center for Education Statistics. They define underrepresented minorities as US citizens or permanent residents who identify as Black or African American, Hispanic or Latino, Native Hawaiian or Other Pacific Islander, and American Indian or Alaska Native.

In 2018, members of these racial groups made up 36.52\% of the graduate-age population, age 25-29. They were in the minority, and they were also underrepresented in physics. According to the latest data, in the United States in 2018 \cite{APS_URM_stats}:
\begin{itemize}
\itemsep0em 
\item 11.9\% of physics bachelor’s degrees went to underrepresented minorities. 
\item 10.1\% of physics master’s degrees went to underrepresented minorities. 
\item 6.7\% of physics doctoral degrees went to underrepresented minorities.
\end{itemize}

Each of these percentages was down from the previous year, but there has not been much variation. Between 2014-2018, the average percentage of bachelor’s degrees in physics that went to URM students each year was 13\%. The average percentage of master’s degrees was 10\%. The average percentage of doctoral degrees was 7\%. This percentage has remained between 3.3-7\% since 1998, demonstrating that generational change without targeted action is an ineffectual solution. 

Over five years, between 2014-2018, the average number of students of different ethnicities earning bachelor’s degrees in physics in the United States each year was:
\begin{itemize}
\itemsep0em 
\item 23.6 American Indian or Alaska Native students--and 76.7\% of bachelor’s degrees awarded to American Indian, Alaska Native, Native Hawaiian or Other Pacific Islander students went to men, 23.3\% to women.
\item 596.2 Asian students--and 77.3\% of those degrees went to men, 22.7\% to women.
\item 236.2 Black or African American students--and 74.5\% of those degrees went to men, 25.5\% to women.
\item 726.6 Hispanic or Latino students--and 80.7\% of those degrees went to men, 19.3\% to women.
\item 9 Native Hawaiian or Pacific Islander students (This group was categorized with American Indian or Alaska Native students in the breakdown by gender.)
\item 5691.8 White students--and 80.9\% of those degrees went to men, 19.1\% to women.
\end{itemize}

Over five years, between 2014-2018, the average number of students of different ethnicities earning doctoral degrees in physics in the United States each year was:
\begin{itemize}
\itemsep0em 
\item 2.4 American Indian or Alaska Native students (0.2\% of all students earning doctoral degrees in physics)
\item 88.8 Asian students (9\% of all students earning doctoral degrees in physics)
\item 19.8 Black or African American students (2\% of all students earning doctoral degrees in physics)
\item 52.2 Hispanic or Latino students (5\% of all students earning doctoral degrees in physics)
\item 1.6 Native Hawaiian or Pacific Islander students (0.2\% of all students earning doctoral degrees in physics)
\item 813.6 White students (83\% of all students earning doctoral degrees in physics)
\end{itemize}

Over five years, between 2014-2018, Native American, Black and Hispanic students were underrepresented among those who earned bachelor’s and doctoral degrees in physics in the United States. Despite making up 1.51\% of the college-age population, Native American students earned an average of 0.4\% of bachelor’s and doctoral degrees in physics. Despite making up 15.64\% of the college-age population, Black students earned an average of 3\% of bachelor’s and 1.8\% of doctoral degrees in physics. Despite making up 21.7\% of the college-age population, Hispanic students earned an average of 9.2\% of bachelor’s and 4.8\% of doctoral degrees in physics.

Black and Hispanic students have consistently earned a smaller percentage of bachelor’s degrees in physics than in computer science, chemistry, engineering, and math and statistics. This strongly implies that physics is uniquely inaccessible to a large swath of people capable of excellence in STEM, which means that the field is losing out on their collective talents due to ingrained biases. 

The US national labs publish demographic data \cite{nat_labs_diversity} about women, underrepresented minorities and other people of color in senior leadership, research/technical management, operations management, technical research staff, operations support staff, postdoctoral positions, graduate student positions and undergraduate student positions. According to their reporting, URMs are only slightly underrepresented among their undergraduate population. URMs steadily decrease in representation as their careers progress from graduate students to staff and research positions to management, with a noticeable drop in representation among postdocs. This state of representation is a significant bottleneck to the future careers of URM scientists. Additionally, the data provided by US national labs combines with all other Asian scientists the groups of Hawaiian and Pacific Islander scientists, who tend to be underrepresented. This combination obscures the underlying demographic data. 

\paragraph{Data on Gender:}

According to APS and AIP \cite{APS_women_stats}, in 2018, women were awarded 22\% of bachelor’s degrees in physics in the US. This was the second-highest percentage of physics bachelor’s degrees that have gone to women in the US since 2002 and 2004, when that percentage was 23\%. This is below the average for all STEM bachelor’s degrees, 37\% of which went to women in 2018. 

In 2018, women were also awarded 22\% of all PhDs in physics in the US. This was the highest percentage of physics doctoral degrees that have gone to women in the US. This is below average for all PhDs, 50\% of which went to women in 2018.

According to a report by the National Science Foundation \cite{hamrick2021}: “Physics has the lowest share of female degree recipients within the broad field of physical sciences.” 

Women are underrepresented in physics education at all levels, and their proportional representation has not increased significantly since the early 2000s for bachelor's degrees and the late 2000s for PhDs. A majority of US institutions accept only 5 students to their undergraduate and PhD physics programs per year. It is possible that the representation of women earning bachelor's and PhDs in physics has stalled at 20-25\% because a large number of institutions reserve a single space, and no more, for female students in their programs per year. 

\paragraph{Data on Sexual Orientation \& Gender Identity (SOGI):}

There is no comprehensive demographic data available to confirm the number of LGBTQIA+ scientists or science students \cite{LGBTclimate}. This is a base measure of whether these scientists and students are being excluded for reasons independent of their expertise and demonstrated capacity for scientific excellence, so it is detrimental to efforts to support LGBTQIA+ inclusion and equity not to collect this data.

Some organizations such as lgbt+physicists\footnote{\href{https://lgbtphysicists.org}{https://lgbtphysicists.org}} maintain individual listings of physicists who choose to publicly identify themselves as sexual and/or gender minority scientists, and these efforts are a great support for the community. But these efforts are not comparable to demographic accounting on the scale necessary to measure and monitor inclusion.

The National Science Foundation compiles one of the most complete demographic data sets of scientists in their Survey of Earned Doctorates\footnote{\href{https://www.nsf.gov/statistics/srvydoctorates}{https://www.nsf.gov/statistics/srvydoctorates}}. For over seven decades, they have asked recent PhD graduates for demographic information, alongside questions about their career paths. They do not currently ask about sexual orientation and gender identity when collecting demographic data. However, there have been strong requests from the community and organizations that represent scientists to include SOGI questions to better facilitate advocacy for equity and inclusion \cite{LGBTscienceMag}. 
There have been calls for the Office of Management and Budget (OMB) to require the NSF National Survey of College Graduates to include SOGI demographic questions. 

Despite the lack of complete or representative demographic information from surveys of the National Center for Science \& Engineering on sexual and gender minorities, there is substantial evidence of LGBTQIA+ underrepresentation within STEM at a rate of about 17-21\% \cite{Freeman_letter}. As with other minoritized groups, LGBTQIA+ individuals are less likely to pursue scientific degrees, or to remain in scientific degree programs once they have started, than the general population as a whole. Adverse and unsupportive environments contribute significantly to the marginalization and attrition of LGBTQIA+ students and scientists \cite{Freeman_letter}.

In April 2021, NSF began a pilot program of including, for the first time, SOGI questions in their surveys \cite{freeman2021stem}.

\paragraph{Data on Disabilities:}

Students with disabilities received 6\% of doctoral degrees in physics and astronomy in 2017. This group earned 7\% of all doctoral degrees in science \& engineering \cite{hamrick_disabilities2019}. People with disabilities represented about 6\% of the US population aged 18-34 \cite{UScensus}. 7\% of employed scientists and engineers have disabilities. And 5\% of all physicists and astronomers have disabilities~\cite{NSFdata}. 

While the fraction of degrees earned and employment of individuals with disabilities seems to match the expectation in the 18--34 age range, data from the same survey indicates that the age of enrollment in higher education is significantly higher for individuals with disabilities (see \cite{UScensus}, Table 2.9). Notably, in the age bracket above the age of 34, the fraction of individuals with disabilities in the U.S. can be as high as 13\%.


\subsection{Current Experiences of Marginalized Individuals in Science} 

While it is not possible to generalize the experience of all marginalized individuals, there are a multitude of ways in which marginalization is experienced by them. Some of these experiences are characterized by: access barriers, uneven representation, power imbalance, selective unfairness, lack of belonging, unequal access or treatment, selective judgement, lack of proper support and mentorship, forms of overt or covert oppression.  All of these factors impact individuals both in their ability to perform their duties, and in their experience both socially and professionally, as well as their mental health. 	

The following examples offer a glimpse at the broad experiences of marginalized individuals in STEM: Despite increased representation of women in STEM in the past decade, 62\% of women in STEM with postgraduate degrees report experiencing gender discrimination at work, and 35\% of them report that their gender makes it harder for them to succeed at work. In male-dominated fields, these numbers go up to 78\% and 48\%, respectively~\cite{E1}.  Roughly six out of ten Black STEM employees in the United States say they have experienced racial or ethnic discrimination at work, and 57\% say their workplaces do not pay enough attention to racial and ethnic diversity~\cite{E2}. Standardized tests, which are known to be poor predictors of academic success in physics Ph.D. programs~\cite{E3}, continue to be used in admissions processes despite the fact that they discriminate against students with disabilities and those who are from historically marginalized backgrounds.

Apriel K. Hodari, Shayna B. Krammes, Chanda Prescod-Weinstein, Brian D. Nord, Jessica N. Esquivel and Kétévi A. Assamagan have written a valuable series of papers we recommend members of the particle physics community read closely to better understand current conditions \cite{hodari2022read,hodari2022power,hodari2022informal,hodari2022policing}. 

In ``Power Dynamics in Physics,'' \cite{hodari2022power} the authors “describe how unfair power dynamics related to various aspects of identity—race, gender identity, gender expression, sexual orientation, and ability status—operate in physics settings.” The paper describes “how those with privilege live comfortably alongside the oppression of others,” and how this lack of awareness or attention, or a focus on perception versus reality, contributes to that oppression. The paper points out that in academia, power dynamics are complicated: “The people at the top don’t have all the power; the people at the bottom are not powerless. The key is figuring out which kinds of power reside where, because power in many cases is more about a web than a line.” 

In ``Informal Socialization in Physics Training,'' \cite{hodari2022informal} the authors write that “claims of objectivity in physics are more of a dream than a reality.” They describe how uneven experiences with the culture of physics, the support of mentors and networks, and the impedance of barriers and exclusionary practices influence the experiences of graduate students and postdocs with marginalized identities in physics. The paper suggests actions individuals and organizations can take to lower barriers for Black, Indigenous, and people of color (BIPOC) scientists navigating this critical time in their careers.

And in ``Policing and Gate-keeping in STEM,'' \cite{hodari2022policing} the authors point out that having a marginalized identity affects a physicist’s experience of the world, both inside and outside of STEM, but that physicists with marginalized identities are expected to suppress those identities, despite the harm this causes, to be taken seriously in physics. This paper delineates the difference between bodily and psychological dangers BIPOC physicists face and the discomforts white physicists often characterize as unsafe.

In her book ``Lived Experiences of Ableism in Academia''~\cite{E4}, author Nicole Brown discusses the challenges that institutionalized and interpersonal ableism create in the lives of faculty with disabilities, as well as their struggle to cope with microaggressions and systemic barriers to their wellbeing. This work highlights the struggle of faculty with visible disabilities as being `expected to be recipients of professional attention, not professionals themselves'.    In addition, the experiences of those with hidden disabilities feel a constant pressure to mask their conditions and 'struggle to pass for non-disabled rather than disclosing their disability'. This book is a recommended reading for all who wish to better understand the effects of ableism in academia. 


\subsection{Marginalization in Collaborations\label{collabintro}}
Research in particle physics and cosmology is often conducted by large collaborations, which can be cross-institutional and international. For example, approximately 3,000 scientific authors are listed on publications from each of the two largest particle physics collaborations associated with the Large Hadron Collider, ATLAS\footnote{\href{https://atlas.cern/discover/collaboration}{https://atlas.cern/discover/collaboration}} and CMS\footnote{\href{https://cms.cern/collaboration/people-statistics}{https://cms.cern/collaboration/people-statistics}}. The governance and structure of these collaborations can differ from the governance and structure of institutions (i.e., universities, research labs, etc.), and the collaboration environment can marginalize individuals in different ways. For example, newcomers may have to on-board and navigate through a large collaboration without much local support (e.g., because their group is smaller and less established or their PI is not directly involved in the practical aspects of their research or not available to represent their interests). Collaboration environments are notably unique in that a very large fraction of interactions happen not in-person but virtually over telecons, messaging applications (e.g., Slack), and email. These spaces can lack the personalized aspects of face-to-face interaction and cause communication challenges. Especially in this environment, a lack of support can exacerbate the isolation of those who already hold one or multiple minoritized identities. In Section~\ref{collabretention}, we describe collaboration-specific inclusion programs.

For a more in-depth discussion of collaboration policies, the ways they currently fail to mitigate harm, and recommendations for how to address these issues, we refer the reader to the ``Climate of the Field Contributed Paper" (\cite{hansensmith_climatecp}, section 4 in particular).

\subsection{Challenges in Collecting Demographic Data for Particle Physics}

Collecting demographic data is important for any organization or institution interested in the recruitment and retention of members of marginalized groups. But there are gaps in this data that make it difficult to demonstrate issues and to determine the effectiveness of efforts to improve. 
In the article ``A push for inclusive data collection in STEM organizations"~\cite{burnett2022push}, researchers found that ``some STEM organizations do not recognize entire groups in STEM, including individuals in sexual minorities (i.e., LGBTQ+ people) or individuals with disabilities." 
Another common issue researchers found was that some demographic groups were lumped together, obscuring information about ``distinctive identities that are frequently relegated to broader demographic classifications," such as those consolidated into the category of ``Asian" or ``Asian American and Pacific Islander." In smaller racial surveys it is also common practice to allow participants to pick multiple answers, but then count those people multiple times or pick one category for them based on the surveyor's bias for the analysis. This can result in inaccurate representations of identities and participants changing their answers based on how they perceive they will be counted. The article points out that, ``[i]n response to survey designs that ignore or obfuscate demographic identities, individuals from underrepresented groups may elect not to respond to certain questions or elect not to complete the survey, introducing a non-response bias into the data. As a result, the collected data do not directly reflect the true demographic composition of the organizations." 

Non-response bias and other forms of information withholding in a survey may also arise when participants perceive that the survey responses may be easily connected to an individual, such as those collected during registration for membership in an organization or in a smaller group. 
The article concludes: ``Cumulatively, these findings suggest that the bulk of professional organizations in STEM are not collecting demographic data that are representative of the true diversity within STEM, which misinforms any subsequent use of the data for supporting or guiding organizational operations such as R\&R."

It is also worth noting that surveying demographics is not the only method to collect information on the status of marginalized groups in particle physics. Qualitative descriptions of experiences, like the writings referenced in Section 2.2, can provide extremely valuable information on the issues impeding the recruitment and retention of members of marginalized groups. In ``How to Promote Diversity and Inclusion in Educational Settings: Behavior Change, Climate Surveys, and Effective Pro-Diversity Initiatives"\citep{moreu2021promote}, encouraging and facilitating discussions among focus groups is suggested as another possible method to gather such qualitative information.

In the case of invisible disabilities, there are additional considerations impairing the collection of complete demographic information. Societal factors can often contribute to both the rates of under-diagnosis as well as the willingness towards disclosure of individuals with hidden disabilities~\cite{E5,E6}. Additionally, given the average differences in the presentations of hidden disabilities by gender, the rate of misdiagnosis for this conditions is significantly higher for women~\cite{E7, E8}.

\subsection{Climate Surveys and Site Visits}

Regularly taking systematic climate surveys that include the collection of demographic data and the collection of qualitative data about the experiences of community members is crucial. Properly publicizing the results of those surveys is also crucial, to allow the community to understand the status quo and  evaluate progress. The University of California, Irvine Office of Inclusive Excellence's data dashboard provides a good example of this type of effort~\cite{uci_oie}.

In addition, conducting external site visits to departments, divisions, and collaborations can be an effective way to learn about the experiences of marginalized individuals and assess the climate of organizations and institutions. The APS Climate Site Visits~\cite{cswp_visit}, overseen by the Committee on the Status of Women in Physics (CSWP) and the Committee on Minorities (COM), are an example of this type of effort. Once approved, APS Climate Site Visits require the participation of faculty, staff, and students. The process includes a two-day visit, followed by a set of written recommendations. The goals of these visits are two-fold: (1) Improve the climate for all, with special attention to women and marginalized groups; (2) Provide assistance to departments to institutionalize positive climate changes. At all stages participants have the opportunity to tap the extensive experience of site visitors for insights on creating a more inclusive, welcoming, and supportive environment. 

%% file: engage.tex
\section{Engaging Marginalized Communities}

One important reason physicists participate in public engagement is to inspire the next generation of STEM professionals. But unless public engagement reaches diverse audiences, recruiting new scientists this way contributes to maintaining a lack of diversity in the field of physics. If designed properly, however, public engagement can contribute to efforts to recruit and retain members of marginalized communities. 

The current HEP community has significant relative privilege and power in the design and implementation of initiatives at the heart of this type of public engagement. For equitable and inclusive access and success for marginalized communities in HEP, the current generation of particle physicists has a duty to embark on a full-scale campaign of engagement with members of marginalized groups. 

\subsection{Key Concepts: Inclusive Science Communication}

Although it is not an innate skill, science communication is not included in the curricula of most science programs~\cite{simis2016lure}. To develop inclusive science communication that effectively invites members of marginalized communities into STEM, scientists must first build their knowledge and abilities in this area. 

Effective communication with members of marginalized groups starts from an understanding of the intersecting types of oppression that those groups face. Effective communication also focuses on interaction and inclusion over the simple dissemination of information~\cite{canfield2020science}. 

One way to work toward this is to move away from the deficit model of science communication, which places scientists on a separate plane from people who have not trained as scientists, assuming total ignorance on the part of non-scientists~\cite{simis2016lure}. Alternative models of communication value the knowledge and expertise that non-scientists already possess, particularly in areas that affect their lives, and recommend scientists focus more on respectful, two-way conversation in their public engagement~\cite{wynne1992misunderstood}.

\subsection{Key Concepts: Public Engagement}

This Snowmass contributed paper purposefully uses the term “public engagement” rather than “outreach” to emphasize a shift toward a more interactive form of science communication, based on current knowledge of best practices. 

The American Association for the Advancement of Science defines “public engagement with science” as “intentional, meaningful interactions that provide opportunities for mutual learning between scientists and the public.” It goes on to clarify that “[m]utual learning refers not just to the acquisition of knowledge, but also to increased familiarity with a breadth of perspectives, frames, and worldviews”~\cite{AAAS_public_engagement}.

To understand the difference between outreach and engagement, it is useful to learn about the related concept of \textit{community} engagement. In “Blog: Defining Community Engagement and Outreach,” Community-Centered Libraries initiative quotes an explanation of the difference between outreach and community engagement by Cindy Fesemyer, director of the Columbus Wisconsin Public Library. She explains: “Community outreach is doing what you do in order to fulfill or advance a specific goal of [an organization]. Community engagement is doing what you do in order to fulfill or advance a specific goal of the community”~\cite{libraries_blog_engagement}.

The Community Engagement Assessment Tool, created by Building the Field of Community Engagement, is also a useful resource. The tool consists of a series of questions to help organizations evaluate whether they are using a framework of outreach or engagement in different areas of their work. It was not created specifically to evaluate public engagement with science, but its explanations of the difference between outreach and community engagement are instructive.

For example, in outreach, “[r]elationships are primarily TRANSACTIONAL, for the purpose of completing a project.” They are “often NOT INCLUSIVE of all racial or cultural groups in the community.” They “can be LIMITED to a few community members, often giving influence to those with the loudest voices.” And they are “SHORT-TERM, so staff have to rebuild them as other projects or issues come up.” In contrast, in community engagement, “[r]elationships are FOUNDATIONAL, continually built between and among people and groups.” They “reflect the DIVERSITY within the community.” They “are built not just with current leaders, but also with people with an interest and/or POTENTIAL TO BE LEADERS.” And they “are transformational and LONG-TERM, so community leaders/members can engage in projects and issues as they come up”~\cite{outreach_engagement_tool}.

Public engagement designed to reach marginalized audiences requires planning and sustained effort. But scientists face barriers to participation in public engagement. The Snowmass2021 contributed paper “The need for structural changes to create impactful public engagement in US particle physics” \cite{adikle2022need} details structural changes that could better enable scientists to participate. These structural changes could also be applied to support any work related to diversity, equity, inclusion, and accessibility.

\subsection{Key Concepts: Cultural Competence}

Another important concept in designing public engagement activities to reach marginalized audiences is cultural competence. 

An individual's culture affects multiple areas of their life, including how they obtain knowledge. When scientific knowledge is shared via methods that privilege a certain culture over others, it can discourage or prevent people who do not come from that culture from accessing the knowledge~\cite{overall2009cultural}.

The Race \& Social Justice Initiative at the Seattle Office for Civil Rights put together a useful rubric to explain to institutions the steps on a ladder from a starting point of ``cultural destructiveness” up to a final goal of ``cultural competence.”

For example, the tool discusses the issue of power dynamics. In a state of cultural destructiveness, ``[a]ccess and power are only given to a privilege[d] group” and “other members are purposely excluded.” In the next step, a state of cultural incapacity, ``[e]ducation is still designed for [the] privilege[d] group and no accommodation is made t[o] try to include other groups.” In the next step, a state of cultural blindness, there is ``[n]o acknowledgement of power differences” and ``power is still held by [the] dominant group.” In the penultimate step, a state of cultural pre-competence, ``[p]ower differences are acknowledged, with some understanding but reliance on others (‘experts’).” In the final step, a state of cultural competence, the ``[t]arget community has a role (real power) in education design and application”~\cite{Seattle_inclusive_outreach}.

Researcher Emily Dawson’s paper ``'Not Designed for Us’: How Science Museums and Science Centers Socially Exclude Low-Income, Minority Ethnic Groups” \cite{dawson2014not} is useful in explaining some of the ways public engagement can fall short due to a lack of cultural competence. The paper provides a nuanced exploration of multiple factors that can prevent individuals from benefiting from informal science education. 

Dawson states that there has been a focus on “barriers” to informal science education, including ``structural barriers of cost or geographic distance and attitudinal barriers such as lack of interest in science.” However, addressing a single barrier is not necessarily sufficient to make an informal science education environment welcoming and accessible to all. Dawson gives the example of museums in the UK, which in the 1990s attempted to alleviate the barrier of cost by providing free entry. As a result, ``[v]isitor numbers rose dramatically. Analysis of who these visitors were suggests, however, that the removal of entrance fees did little to diversify museum visitors; existing visitors simply visited more often.” 

Dawson points out that ``[t]he ‘barriers’ approach has been criticized as assimilationist for requiring participants to change to fit institutions, privileging dominant knowledge and practices, and pathologizing others.” Dawson’s paper, which the authors of this paper recommend to those interested in reaching new audiences via public engagement, “provides concrete examples of social exclusion from [informal science education] and suggests that understanding experiences of difference, discomfort, and inaccessibility are crucial for creating more inclusive… practices.” 

As Dawson's paper illustrates, it is essential to partner with individuals in a community to effectively plan public engagement activities with and for that community. This process can reveal hidden barriers and reveal the way that multiple barriers interact. Partnering with members of communities can ensure that public engagement is designed in a way that effectively values and connects with a community and allows individuals to fully benefit from their participation.

\subsection{Checklist: Planning Public Engagement Events for Marginalized Communities}

Engaging marginalized communities is a priority for many scientific institutions. Creating programs specifically designed for members of those communities can be an effective way to start building relationships and encourage participation in STEM.

The following questions are intended to help institutions plan such events and begin to shift their thinking toward a model of community engagement (see other tools above). It is important to consider the audience; to identify and remove barriers; to value partnerships; and to build lasting relationships by thinking long-term and considering the bigger picture.

\begin{itemize}
\item \textbf{Consider the audience:}
\begin{itemize}
\item Who specifically are we hoping to reach with this event? Why are we hoping to reach these communities?
\item How can we plan this event to make it maximally beneficial to these communities? What elements of this plan can we continue to use in other events?
\item What are the best ways to communicate about this event with members of these communities? Can we continue going to those same channels to communicate about other events?
\item Have we created a process by which we take time to evaluate the success of the event after it concludes? What metrics (both qualitative and quantitative) will we use? Which of these metrics will we continue to use in evaluating other events? 
\end{itemize}

\item \textbf{Identify and remove barriers:}
\begin{itemize}
\item Are there logistical barriers (e.g., time of day, day of the week, public transportation access, affordability, safety concerns, financial barriers) to our events that make them inaccessible to these communities? What will we do to address these barriers?
\item Have we allowed adequate lead time and budget to make this event accessible to all members of these communities, including those with disabilities? Have we identified partnership or staffing needs required to make the event accessible?
\end{itemize}

\item \textbf{Value partnerships:}
\begin{itemize}
\item What members of these communities will make good partners in this event? Have we made sure they’re involved in planning the event? Have we secured an adequate budget to support fair compensation for our partners as co-creators of the event, prior to requesting their labor? 
\item Do any members of these communities work for our institution? If they do, do they work in roles with decision-making power (e.g., managerial positions), or do they work primarily in service roles? If members of these communities do not work at our institution, or work only in lower-level positions, is our institution making any effort to change this? 
\item Are members of these communities who work for our institution participating in this event? If so, are they receiving the support they need to take on this effort and fulfill their other job duties? Do they have decision-making power over the planning and execution of the event? Are they being fairly compensated and recognized for their efforts? 
\end{itemize}

\item \textbf{Build lasting relationships:}
\begin{itemize}
\item Is this event a part of a larger effort to build relationships with members of these communities? If so, what is the long-term plan? Who will be responsible for enacting it?
\item Are there ways in which our institution is causing harm to members of these communities? If so, how is our organization working to change this?
\item How are representatives of our institution involved in these communities outside of this event? Are there ways our institution can work with members of these communities on their priorities, even ones that do not directly benefit our institution?
\end{itemize}
\end{itemize}

%% file: pathway.tex
\section{Infrastructure - Pathway, Recruitment, and Retention}

\subsection{High School Engagement}
\label{A}

High school is often considered a crucial moment at which to foster an interest in physics or other related STEM discipline. Especially by junior or senior year, students are already applying for college and thinking critically about possible careers. To increase the representation of marginalized communities in the field of physics, we must make sure that marginalized groups have adequate and equitable opportunities to be exposed to such disciplines. 

One way to ensure equitable access and success is to support existing initiatives that engage and mentor high school students from marginalized communities. Several programs for high school students across the US promote diversity and inclusion in STEM. Many specialize in exposing students to active areas of physics research. A sampling of these are listed below, with links to their websites for more information. This is not an exhaustive set of programs by any means, but it does provide a starting point from which to expand or create new programs to provide opportunities to students in marginalized communities.

\begin{itemize}

\item \textbf{STEP UP} is a national consortium of physics teachers, researchers, and professional societies that have developed and implemented high school physics lessons aimed at encouraging young women to consider pursuing physics \cite{STEP_UP}.

\item \textbf{TARGET} is a 6-week summer internship opportunity for Illinois high school sophomores and juniors interested in physics or other STEM-related fields. It seeks to increase the representation of underrepresented minorities (Black, Hispanic/Latino, Hawaiian/Pacific Islander, Alaska Native/American Indian students) and women in STEM. The interns in this program have the opportunity to work alongside scientists and engineers at Fermi National Accelerator Laboratory \cite{TARGET}.

\item \textbf{ACT-SO}, run through Argonne National Laboratory, is a STEM enrichment and mentoring program that provides research experiences to African American high school students. Since 2013, it has partnered with Fermi National Accelerator Laboratory \cite{ACT_SO}.

\item \textbf{Space Explorers} is a program from the Kavli Institute for Cosmological Physics at the University of Chicago. The multi-year program offers youth from around Chicago exposure to research and the culture of science. Most participants are African American, and more than half are women  \cite{kicp_education_and_outreach_space_explorers}.

\item \textbf{STEM Starters} is a program run by Columbia University graduate students. It is a monthly opportunity for middle school and high school minority students to explore various STEM fields. The program has recently provided several virtual options for continued STEM enrichment \cite{Stem_starters}. 

\item \textbf{COSMOS} is a learning platform designed and sponsored by New York University for K-12 students in New York City, particularly West Harlem \cite{COSMOS}.

\item Stanford University has several high school engagement programs, including: \textbf{Advanced Science Exploratory Program} (seminar series), \textbf{Splash!} (a set of high school exploratory classes on many topics including STEM), and \textbf{“Girls Teaching Girls to Code.”} All of the programs seek to increase diversity and inclusion in STEM-related fields \cite{Stanford}.

\end{itemize}

While this list shows that several programs for high school students do exist, we recommend creating integrated support systems that will keep students continuously involved in the particle physics community from high school to grad school and beyond, to address the ``leaky pipeline" that fails to carry members of marginalized communities from high school through careers in HEP \cite{staff_sousa_2020}.

Rather than forming loose connections with students from marginalized communities via one-time speaking engagements or one-day school trips, we recommend researchers engage these students year-round. We envision an ``adopt a school" or ``adopt a school district" program that researchers can sign up for, with funding provided by DOE or NSF. One could even create a consortium of area-wide or county-wide scientists who could make regular visits to schools, offering students a variety of perspectives.

Having access and opportunities for students, particularly those from marginalized communities, is important at all levels of education. In addition to supporting these programs at the high school level, the physics community should also support programs for students in middle school and elementary school. As a long-term goal, we would like to have a more unified list of programs that focus on younger students as well.


\subsection {Augmented REUs and Research at Community Colleges}
\label{B}

A good way to instill a sense of belonging in STEM in students is to involve them in research. The NSF-funded Research Experiences for Undergraduates (REU) programs and other equivalent research programs for undergraduate students are an excellent way to get students involved in research early on. 

Specific efforts to encourage the participation of students from marginalized groups are important. One example is the LAMAT Institute at UC Santa Cruz, an 8-week summer REU program. Two-thirds of the program's students come from historically marginalized groups in STEM, and two-thirds are first-generation college students~\cite{Lamat}. Roughly 74\% of the program's alumni have gone on to attend graduate school; 16 of the students have become NSF graduate fellows, and 3 have become NASA Hubble Postdoctoral Fellows \cite{Lamat}. 

A good way to support students from marginalized communities is to provide research opportunities to students at 2-year community colleges. Many economically disadvantaged and first-generation college students take advantage of community colleges, which offer a cost-efficient path toward a college degree and, in many cases, provide extra resources to help students adapt to college life. 

While some US community colleges have pipeline programs in place to transfer students to 4-year universities (sometimes with R1 status) to continue their studies, there are few well-defined programs available to give research opportunities to students at community colleges prior to matriculation. To bring more students from marginalized communities into the particle physics pipeline, we recommend supporting more programs that provide research experiences to these students. One example of such program is the STEM Summer Research Opportunity (SRO) Program~\cite{SRO} at Chaffey College in Southern California.

\subsection{Bridge Programs}
\label{C}
Bridge Programs at universities across the US are another way to increase and promote diversity in physics \cite{doi:10.1063/PT.3.2511}. They provide students with mentors, research experience, and other opportunities to help them succeed in a physics PhD program \cite{doi:10.1063/PT.3.2511}. One example is the American Physical Society's Bridge Program~\cite{APS_bridge}, which has supported URM students' pursuit of graduate programs at the program's partner sites around the country, including the five original bridge sites - Florida State University, California State University Long Beach, Indiana University, The Ohio State University, and University of Central Florida.

Additional bridge programs exist, and still others are being developed. The programs report impressive retention rates, surpassing the national average of roughly 60\% of graduate students who stay in their PhD programs. A 2017 paper reported a retention rate of 80\% for the Fisk-Vanderbilt bridge program and of 90\% for the University of Michigan's Imes-Moore bridge program in applied physics~\cite{beckford2017survey}. Another example is the Cal-Bridge Program~\cite{Cal-Bridge}, a collaboration between California State University campuses, whose majority of students are from marginalized backgrounds, and University of California campuses. The Cal-Bridge Program includes mentoring and career development, in addition to summer research opportunities via the ``Cal-Bridge Summer (CAMPARE)'' component.   

It is important that the physics community gain a better understanding of the effectiveness of bridge programs like these. We recommend gathering additional statistics, including those related to student outcomes, and pursuing qualitative studies, including those related to student experiences. 

\subsection{Programs for Veterans and Other Non-traditional Students} Very few programs provide pathways into physics degree programs or the physics workforce for veterans or other non-traditional students. One example is the Fermilab VetTech Internship Program~\cite{vettech}, through which veterans intern as technicians and contribute to Fermilab experiments, using the technical skills acquired in their military and academic experiences. It is important to create and support programs like this one.

\subsection{Programs for Refugees and Undocumented Individuals}

Refugees and undocumented individuals face unique challenges in accessing education and employment opportunities. It is important for institutions to develop programs to support members of these marginalized groups.

The REFUGES program~\cite{refugees}, launched in 2012 at the University of Utah, offers a platform to support marginalized communities. The program's objective is to provide training in STEM for refugees, nonnative English speakers, economically disadvantaged students, and other marginalized students. The program endeavors to address challenges that refugee youth face, such as being placed in school grades inconsistent with their level of education. The REFUGES program is built on concerted efforts between the academic community of the University of Utah and local church communities, parent groups, and social organizations. This program may serve as an example of how the particle physics community can promote a culture of  equitable access.

\subsection{Identity-based Conferences and Organizations}

Identity-based conferences and organizations can provide safe spaces for people from marginalized groups to access support networks while pursuing professional experiences and developing career skills. Theses conferences and organizations have played a crucial role in the retention of members of marginalized communities in physics. For example, it has been shown that the American Physical Society Conference for Undergraduate Women in Physics (CUWiP)~\cite{CUWiP} has contributed to increasing the number of women pursuing PhDs in physics. Existing identity-based conferences and organizations include:
\begin{itemize}
\itemsep0em 
\item American Physical Society Conference for Undergraduate Women in Physics (CUWiP) \cite{CUWiP}; 
\item American Physical Society National Mentoring Conference (NMC)~\cite{NMC}; 
\item Black In Physics Week~\cite{BiP}; 
\item National Society of Black Physicists (NSBP)~\cite{NSBP}; 
\item National Society of Hispanic Physicists (NSHP)~\cite{NSHP}; 
\item Out in STEM (oSTEM)~\cite{oSTEM}; 
\item Society for the Advancement of Chicanos and Native Americans in the Sciences (SACNAS)~\cite{SACNAS};
\item Society of Indigenous Physicists~\cite{SIP}; 
\item The SciAccess Initiative~\cite{SciAccess};
\item VanguardSTEM~\cite{vanguard}.
\end{itemize}
One pitfall of identity-based conferences and organizations is that most focus on a single identity and may fail to recognize issues related to intersectionality~\cite{yabasic}.

\subsection{Effectiveness of Inclusion Programs} \label{D}

To evaluate the effectiveness of programs such as these, the physics community needs metrics, such as the percentage of participants who remain in STEM degree programs and the percentage who go on to STEM-related careers, along with qualitative measures of student experiences.

At this time, we find it difficult to obtain metrics by which to judge the effectiveness of high school programs. 

Some inclusion programs at the undergraduate level have provided useful data. One such program is MemphiSTEP, an NSF-funded program at the University of Memphis \cite{windsor_bargagliotti_best_franceschetti_haddock_ivey_russomanno_2015}. The five-year program was designed to boost the number of STEM graduates via support and readiness programs such as a Math Bootcamp, learning communities, networking opportunities, and research experiences. The program tracked progress in GPA and STEM retention rates among graduates and found, among other things, that the retention of STEM majors increased by 17\% during the course of the program. The average retention rate for Black freshman STEM majors increased from 44 to 81\% \cite{windsor_bargagliotti_best_franceschetti_haddock_ivey_russomanno_2015}. 
 
Another program for undergraduates is the Meyerhoff Scholars Program, based at the University of Maryland - Baltimore County. Between 1989 and the summer of 2018, 70.8\% of STEM undergraduates enrolled in the program were classified as URMs. Most of those participants earned their bachelor's degrees in science or engineering, and 75.8\% of those students continued on to graduate or professional programs. Students who participated in the program were twice as likely as those who declined to earn a degree in a STEM field, and they were five times as likely to continue on to graduate school \cite{doi:10.1126/science.aar5540}.  

We recommend more programs for high school and undergraduate students, especially those sponsored by NSF and the US Department of Energy's Office of Science, publicly release the statistics that can be used to gauge their effectiveness. 

\subsection{Flexibility and Support for Individuals with Disabilities and for Neurodiverse Individuals}

Members of the particle physics community also experience marginalization related to visible and invisible disabilities, along with differences in neurocognitive functioning. Findings and recommendations from a community survey about disabilities and mental health are detailed in a 2022 white paper~\cite{Assamagan:2022oka}. We recommend additionally considering the neurodiversity of the particle physics community to ensure that all members are able to participate to the fullest extent.

Programs supporting individuals with disabilities are common in institutions of research and higher education. Typically, these programs ensure compliance with the Americans with Disabilities Act~\cite{ada} by ensuring access to reasonable accommodations for a given disability and safeguarding the status of individuals with disabilities as a protected class. Due to the rapidly evolving field of neurodiversity research, as well as the complexity of this area, many of the disability services offered by these programs can be insufficient or inappropriate for the needs of all disabled individuals. For example, one common accommodation available for neurodiverse students at most institutions is providing additional time for completing tests and assignments. However, studies have found that the impact of such accommodations on student learning is minimal, and their impact on student performance is far from sufficient~\cite{ada2}. Research-based programs specifically targeting students with neurocognitive differences are essential to provide them with equitable access to educational resources and fair evaluations.


The challenges specific to individuals on the autism spectrum and those with ADHD are useful to exemplify the benefits of flexibility and support in the workplace, due to the prevalence and extensive body of research for both conditions. Given that these hidden disabilities often present themselves as behavioral traits i.e. poor time management or social abilities as well as positive qualities such as increased focus or divergent thinking, flexibility in both the environment and expectations are crucial to enable their success. 

Research suggests that neurodiverse  individuals can make a significant impact in terms of creativity and innovation when provided with an environment that potentiates their capabilities. However, these conditions can have slightly different presentations which vary by individual, and often present comorbidity with other hidden disabilities as well as mental health conditions. An environment with increased flexibility of work locations, schedule, and adaptability of the available accommodations have been observed to generate positive impact to both the individuals and the productivity of the workplaces globally.

%% file: culture.tex
\section{Culture Shift - Creating an Environment Conducive to Access and Success for Marginalized Communities in HEP}

\subsection{Addressing Misconduct Including Harassment, Discrimination, and Bullying\label{misconduct}}

\input{misconducts}
    
\subsection{Improving Daily Experience\label{improvedaily}}
\input{daily_interactions}

\subsection{Collaborations\label{collabretention}}
As mentioned in Section~\ref{collabintro}, much of the research in particle physics and cosmology is conducted in large collaborations. Here, we focus on the issues marginalized individuals face in collaboration environments.

\begin{itemize}

\item Just as individual institutions need clear sets of governance policies (including values statements, codes of conduct, and descriptions of accountability structure), so too do scientific collaborations. As evidenced by the recent Petition for Codes of Conduct in Astronomy Collaborations \cite{codeofconducttweet}, many members of the astronomy community endorse the importance of collaboration codes of conduct. Even when codes of conduct are in place, individuals from marginalized communities often do not report breaches, due to the emotional toll and time away from research it takes. Coming forward rarely results in action to address the problem and may lead to retaliation. For this reason, codes of conduct must be accompanied by clear accountability structures and procedures. As suggested in Section~\ref{misconduct}, collaboration members should have access to multiple supportive structures for reporting, such as Ombuds and anonymous feedback forms seen only by designated collaboration leaders.

\item Individual collaboration members can contribute to improving the daily experiences of their marginalized colleagues (see Section~\ref{improvedaily}) by educating themselves about microaggressions, particularly those that take place where most collaboration interactions occur (over telecons, messaging applications, emails) \cite{sue2019disarming,sue2020microintervention}. Collaborations can support this by providing training. 

\item Collaborations can improve the daily experiences of individuals from marginalized communities by providing accessible training, mentoring, and support networks \textit{specific to minoritized collaboration members}. Student experiences with  racial and ethnic discrimination in their interactions with faculty is linked to negative STEM retention outcomes \cite{park2020student}, so those responsible for mentoring and teaching require training as well. 

\item A study~\cite{teamup} produced by the TEAM-UP Task Force recommends collaborations provide support to relieve financial stresses and burdens on Black undergraduate students. We recommend collaborations provide funding to support members coming from marginalized backgrounds.
\end{itemize}

\subsection{Enabling Pathways in Academia, Industry, and Beyond}

Training in physics can enable careers in a variety of areas such as industry, non-profit organizations, government, and academia \cite{mulvey2020new}. However, the faculty and university staff responsible for training students and early-career researchers typically lack both experience with and knowledge of non-academic career paths. Leaving academia is stigmatized in academic environments, as faculty advisors explicitly or implicitly place lesser value on alternative career paths \cite{frank2019commentary, teamup,national2021pathways}. These negative attitudes marginalize students and researchers who veer from the traditionally accepted progression through academia (i.e., earning degrees and seeking academic positions, with minimal time spent outside of academia) and compound challenges faced by those who already hold other marginalized identities. Presenting a narrow academic pipeline as the single path to success is exploitative, given the relative lack of long-term or permanent jobs in academia \cite{teamup,national2021pathways}.

A best-practices guide for federal agencies put together by the National Science and Technology Council lists this topic, described as ``STEM Pathways," among four ``Key Areas for Advancing Diversity and Inclusion in STEM"\cite{nstc_bestpractices_stem2021}. Reports more specific to physics and astronomy, such as those by the Joint Task Force on Undergraduate Physics Programs \cite{mcneil2017preparing}, TEAM-UP \cite{teamup}, and the 2020 Astronomy Decadal Survey \cite{national2021pathways} contain highly relevant discussions of the topic as well. For example, a rubric in the TEAM-UP report \cite{teamup}, which can be used to determine how far a physics department has progressed toward implementing the goals of the report, includes a section related to career options. In the advanced stage, the report says, departmental talks will feature alumni who have gone on to a variety of career paths, inside and outside of academia, about whom faculty will  ``speak with equal pride." Multiple reports recommend supporting students who go on to careers outside of academia by adding technical or computing training to curricula, by strengthening connections with alumni who represent a diversity of career paths, and by changing the attitudes of faculty mentors. A program already putting these recommendations to work is the Erd\H{o}s Institute based at The Ohio State University, which provides a purpose-built seminar series, training, and access to professional networking beyond academia~\cite{erdos}.
HEP has obvious links to industry through instrumentation and big data analysis and should think creatively about how to open pathways that can flow in both directions --- from HEP to technology development for industry, software engineering, data science, etc., and back into HEP. Programs across all levels, from pre-baccalaureate to graduate and even postdoctoral, could help cast a wide net to recruit and support for these pathways members of marginalized communities, including those considered to be on non-traditional trajectories (which include gap years, military service, industry, time spent to care for families, and more).

\input{promotion_hiring}


%% file: misconducts.tex
Harassment, discrimination, and bullying occur in our community. These behaviors cause severe and often irreparable harm to careers and lives. Despite years of effort, we still have not figured out how to satisfactorily address these issues and prevent future harm. 

It is illegal in the US to discriminate (the definition of discrimination here includes harassment) based on an individual's race, color, national origin, religion, sex, age, disability, sexual orientation, pregnancy, gender identity, and genetic information. It is also illegal to retaliate against someone for opposing or reporting a discriminatory practice, or participating in a complaint process\footnote{\href{https://www.eeoc.gov}{https://www.eeoc.gov}}. In recent years, institutions have provided mechanisms to report and investigate discrimination through Human Resources and Title IX offices, and often, university staff and faculties have mandatory reporting responsibilities regarding illegal discrimination. However, our community's process of addressing abusive and wrongful misconduct remains inadequate.

First, the legal reporting process can further traumatize or otherwise cause additional harm to those who experience discrimination \citep{holland2021reporting}. Retaliation, though prohibited by law, is common\footnote{\href{https://www.eeoc.gov/select-task-force-study-harassment-workplace}{https://www.eeoc.gov/select-task-force-study-harassment-workplace}}. It is understandable, then, that according to the U.S. Equal Employment Opportunity Commission, an estimated 75\% of individuals in the workplace who have experienced sexual harassment have never talked to supervisors or advisors about it. Severe under-reporting occurs in academia as well (see results from a longitudinal study in the field of astronomy in \cite{harassment} and additional studies for academia in  \cite{ cantalupo2018systematic, hardebeck2017report}). 

Experts \citep{holland2021reporting} have emphasized the importance of providing supportive structures for reporting and investigating misconduct, and have emphasized the importance of providing supportive structures independent of the reporting and investigation processes. Recent improvements include the use of anonymous reporting systems and Ombuds\footnote{\href{https://www.ombudsassociation.org}{https://www.ombudsassociation.org}}. Organizations recommend designating faculty/staff who can explain the reporting process, as well as training faculty/staff on how to support those who experience discrimination beyond the reporting process. It is  worth noting that campus counseling and mental wellness centers are often tasked with supporting students and scholars in these situations, but that those centers are often overwhelmed, a problem that may have become more severe during the Covid-19 pandemic \citep[see discussions in][]{brown2020did}.

Not all abusive misconduct can be addressed by current reporting processes. There may not be a channel to address misconduct that is prohibited by institutional or community standards, but that falls below the threshold of a legal violation. One example is bullying, which usually refers to ``a sustained display of hostile verbal and nonverbal behaviors, excluding physical contact". Examples of bullying in academia include public humiliation and threats to an individual's project or visa \citep[See examples in][]{academic_bullying_stories}. Bullying can also take place on the internet \citep{tokunaga2010following, noakes2021distinguishing}. 

It is important for academic institutions to enact formal policies and guidelines, and to provide training to students, faculty and scholars regarding bullying \citep{ghosh2011toxic}. Studies describe bullying as more prevalent in the academic workforce than within the average population\citep{moss2021stem, keashly2010faculty, hollis2012bully}. Although bullying can appear in a wide range of power structures, it often involves power disparities, and it is more likely to affect members of marginalized groups. Perpetrators are more likely to be from highly ranked institutions and positions (including, for example, director of US OSTP\footnote{\href{https://www.politico.com/news/2022/02/07/eric-lander-resigns-00006545}{https://www.politico.com/news/2022/02/07/eric-lander-resigns-00006545}}). Members of sexual and gender minorities \citep{sallee2013sexual, lampman2012women} are more likely to report being bullied. International scholars are more likely to experience a greater severity of bullying \citep{moss2021stem}. It has been proposed that bullying is sometimes employed as a career tool in academia, to maintain power structures and eliminate perceived competitors~\citep{khoo2010academic, tauber2022bullying}. In those cases, proving violations is often difficult. Victims may be told that ``nothing can be done" if a case doesn't involve illegal discrimination (described above). In those cases, the victim is offered no recourse against retaliation.  

The scientific community faces a unique challenge in addressing cross-institutional misconduct, such as misconduct that occurs during a conference or during an academic visit, including travel to speak at a colloquium or seminar. Scientific collaborations often involve participants from multiple institutions. Failures to address cross-institutional misconduct can cause persistent daily harm (we refer the reader to the ``Climate of the Field Contributed Paper" \cite{hansensmith_climatecp}, section 4, and a contributed Letter of Interest, ``A Need for Alternative Collaborative Means to Address Misconduct" by Kimberly Palladino \citep{Palladino_LoI} for more details). A few considerations include:
\begin{itemize}
\item It is particularly important to recognize and formalize the rules for cross-institutional misconduct. For example, if the misbehavior involves or impacts individuals from different institutions, it should be made clear which institution is responsible for investigation and disciplinary actions. Scientific collaborations should work with institutions to ensure the legal basis of those measures.
\item Funding agencies and professional societies can step in to address cross-institutional issues. The Decadal Survey  by the Astronomy Community recommends \citep{national2021pathways} that “NASA, NSF, DOE, and professional societies should ensure that their scientific integrity policies address harassment and discrimination by individuals as forms of scientific misconduct.”
\item Institutions should also enact policies \citep{gluckman2022, mervis2019universities} to battle the ``pass the harasser" phenomenon \citep{brown2019pass}, in which a professor or administrator is credibly accused of misconduct at one institution, then transitions to a different institution. The first institution remains silent about the misconduct because of a settlement agreement or out of fear of a defamation lawsuit, and the professor or administrator is able to continue the misconduct at the new institution. Both the University of California, Davis (UC Davis) \citep{ucdavis} and the University of Wisconsin System \citep{uwisconsin} have detailed their policies to combat this phenomenon.
\end{itemize}


%% file: daily_interactions.tex
Although more subtle than the types of misconduct mentioned above, daily social interactions can also contribute to adverse experiences of marginalized individuals. As Physicist Dr. Kétévi Assamagan said in an interview with \textit{Physics Today}, ``There were people who didn’t believe in my abilities. The prejudice is not provable, but it’s there in the daily interactions that you don’t feel good about. You feel something is not right”~\citep{physics_today_Assamagan}. Or as mentioned by Computer Scientist and Mathematician Dr. Lenore Blum, ``Subtle biases and micro-aggressions pile up, few of which on their own rise to the level of `let’s take action', but are insidious nonetheless”\citep{lenore_blum_resignation, low_bias_workplace}. These daily indignities can have a great impact on the work experience of members of marginalized groups. For example: 
\begin{itemize}
\item Working remotely during the Covid-19 pandemic, minoritized Black workers in so-called “knowledge” job categories reported better job experiences, despite the drawbacks of working from home, possibly because of the removal of daily tension \citep{ameri2022leveling, black_working_home}. 
\item A recent study \citep{low_bias_workplace, du2021insidious} reveals that a very low level of daily bias against women over an extended period of time can create significant gender disparities in career progression. Brief, powerful interventions such as gender quotas may have temporary effects, but disparities remain when the underlying, persistent bias is not addressed.
\end{itemize}


Studies, as well as lived experiences, have revealed a range of bias patterns marginalized individuals experience in daily interactions, including microaggressions \citep{pierce1970offensive, pierce1977experiment, sue2007racial}, unequal expectations related to service work \citep[excessive service work load expected from, e.g., African American women faculty, ][]{harley2008maids}, stereotype threat \citep{steele1995stereotype}, and imposter syndrome \citep{parkman2016imposter}. Marginalized individuals have described difficulties with informal networking and developing beneficial career connections \citep{physics_today_Lalanne, physics_today_Mtingwa}, which has also been studied in literature \citep{falci2020network, xu2011gender}.
Academic institutions and groups have much to do to ensure a better social environment for marginalized groups, as has been recommended in literature \citep{ali2021actionable, chaudhary2020ten, teamup, ong2018counterspaces, rodriguez2020we, richey2019gender}.  
For example, to counteract microaggressions \cite{sue2019disarming,sue2020microintervention}, institutions should describe a strategic framework that includes concrete steps that bystanders can take (``microinterventions") to move away from passivity and toward disrupting the harm, educating the microaggressor, and supporting the individual harmed. We strongly urge academic institutions and groups to study and follow the guidelines provided in the expert studies cited above.

Cross-institutional and cross-disciplinary efforts have emerged to facilitate learning and to work toward creating a culture of diversity, equity, inclusion, and accessibility in STEM. These efforts include the American Physical Society's APS-IDEA alliance (APS Inclusion, Diversity, and Equity Alliance), the American Association for the Advancement of Science's STEMM Equity Achievement Change (SEA Change) program, and The American Association of Colleges and Universities' Teaching to Increase Diversity and Equity in STEM (TIDES) program. These programs provide training opportunities and supportive structures to help organizations learn and devise strategies for change. Professional societies, including the American Physical Society, have also provided series of online talks to raise awareness or provide training related to DEIA. As recommended by the American Elasmobranch Society \citep{shiffmancan}, professional societies can facilitate change through measures including amplifying marginalized voices in the society's communication channels, providing education opportunities to members, ensuring inclusive meetings and talks, setting up mentoring programs, and hiring an independent safety officer.

As the diversity of our field increases,  so does the urgency for improved resources for education and awareness of the issues and barriers that DEAI must overcome to truly empower and support marginalized individuals. Beyond training and education programs, increased exposure to the specific experiences of the marginalized members of our community is a crucial component to improve our commitment to DEAI efforts as a community. A deep understanding and true empathy for the experiences of the marginalized is essential to build a strong foundation for progress in this area.

%% file: promotion_hiring.tex
\subsection{Evaluation and Selection Practices}

Students and scholars in particle physics face many evaluation and selection processes: student evaluations, manuscript reviews, funding (or other research resources) proposal reviews, school/job applications and job promotions. Unfortunately, many elements of these processes are known to contain biases \cite{Moss-Racusin16474,Eaton2019} that unfairly disadvantage people with marginalized identities. Consider, for example \citep{white2020facade}, that ``fitness" in faculty applications can be used as a tool to exclude people from marginalized backgrounds, or that fewer STEM career ads are shown to women \citep{lambrecht2019algorithmic}. Selection and promotion practices often fail to take into account multiple factors that could bias a process toward a single gender. For example, in a laboratory experiment, researchers found that their subjects, both men and women, were twice as likely to hire a man over a woman to perform a mathematical task. In addition, when they asked job candidates to self-report about their performance of the task, the subjects doing the hiring failed to take into consideration that men, on average, are more likely to overstate their performance, and women are more likely to understate it \citep{reuben2014stereotypes}. 

Our field's selection and promotion practices were designed by the existing workforce, which is not diverse or inclusive. These practices often fail to consider the realities that people with marginalized identities face. It has been noted that gender and/or racial minorities spend significantly more time on service work, which is less valued than other work in selection and promotion processes \citep{harley2008maids}; women in physics, who are more likely than men in physics to be married to a fellow physicist, are more likely to face the ``two-body problem" \citep{mcneil1999dual}; marginalized groups suffer from lack of culturally responsive mentorship, which impedes their access to job and promotion opportunities \citep[see  examples of barriers faced by women of color STEM faculty][]{corneille2019barriers,ong2018counterspaces}; and marginalized groups endure more hostilities in their work environments \citep[e.g., see studies about gender inequality in STEM fields ][]{casad2021gender}. Subtle disadvantages (defined as ``micro disadvantages" in \citep{roos2009gender}) can compound to cause significant harm to marginalized individuals. For example, gendered, racist microaggressions from students can hinder a Black female faculty member's ability to develop a competitive promotion or tenure packet \citep{pittman2010race}. 

Hiring teams' ideas about a woman's role in society may prevent them from making initial offers, or segregate the roles for which offers are made along gendered lines. For example, if a hiring committee assumes women cannot weld or would not be interested in welding, they may refuse to hire female instrumentalists, creating segregation within a department or institution. Some institutions prefer not to hire women because they are perceived as potential mothers, who they assume may prove incapable of caring about their work once they have children. (Potential fathers, it should be noted, are not treated with the same wariness.) Some university departments may also refuse to invest in female students due to similar concerns about their ability to persist in the field \citep[][and references therein]{SASSLER2017192}. When a woman is hired, she may be offered a lower salary.

There have been many recent efforts to standardize and improve the academic evaluation and selection processes, but the successes of those initiatives remain largely unquantified. However, the potential to improve those processes through data collection and close collaboration with domain experts is powerfully illustrated by the example of the Hubble Space Telescope proposal review process.

The Hubble Space Telescope (HST) allocates telescope observation time to members of the astronomy community through a competitive peer-review process. In 2014, studies of the HST proposal selection results revealed a persistent pattern of privileging proposals submitted by male PIs over those submitted by female PIs \citep{Reid_2014}. Measures such as increasing the gender diversity of the selection committee and providing training on unconscious bias did not remove the pattern. So the Space Telescope Science Institute (STScI) collaborated with social scientists, who observed bias in discussions about proposers' qualifications in the peer-review process. After the finding, STScI established a thoughtfully designed, dual-anonymous review process \citep{strolger2019doling, 2019BAAS...51g.272S}, which both significantly alleviated the gender bias and increased the success rate of early-career researchers in HST proposal selections. In 2020, inspired by this success, NASA's Science Mission Directorate also adopted a dual-anonymous peer review (DAPR) process for all Research Opportunities in Space and Earth Science \citep{Berkowitz_duarp_traction}. This success story also provides valuable lessons for other selection processes \citep{2019BAAS...51g.272S, aloisi2019conscious}. 

Although the above example illustrates that efforts must go beyond implicit bias training, we do commend the recent addition of discussions of implicit bias to ad hoc reviewer training and to grant proposal review panel processes. It would be beneficial to build such training into the standard proposal review process and implement this practice uniformly across all funding agencies.    

Another important step toward diversity, equity, inclusion, and accessibility is for the community to elevate the importance of DEIA work in evaluation and reward processes. We consider the requirement to demonstrate contributions and commitment to DEIA through a mandatory ``Statement of Contributions to Inclusive Excellence'' to be a positive development, along with the addition of discussions of DEIA to the interview process in faculty searches at an increasing number of institutions.

%% file: recommendations.tex
\section{Summary of Recommendations}

Based on the findings described in the report, we identify the following general recommendations for the US particle physics community:
\begin{enumerate}
    \item To improve the experiences of members of marginalized communities, engage these communities, collect feedback, assess the effectiveness of existing programs, and develop best practices;
    \item Sustain engagement with marginalized communities, and train members of the particle physics community for productive engagements;
    \item Create infrastructure to improve academic, financial, and personal support for members of marginalized communities;
    \item To create an environment conducive to success and the retention of members of marginalized communities, establish community expectations, foster inclusion, and ensure individual and institutional adherence.
    \item Establish a mechanism to monitor progress in the area of DEIA, including the implementation of recommendations resulting from the Snowmass 2021 process.
\end{enumerate}
 
 Below we summarize specific recommendations for different stakeholders: Funding agencies, professional societies (APS, AIP, AAAS, etc), institutions (universities, national laboratories, etc), scientific collaborations, and individual scientists.

\subsection{Collect Feedback, Develop Best Practices}

\paragraph{Recommendations for Funding Agencies}
\begin{itemize}
\item Collect demographic data of grant recipients and make these data available to help evaluate progress toward equitable access in particle physics. Where funding goes, so should demographic statistics. In particular for project and facility grants, agencies should know what they are paying for, but also who they are paying, and perhaps more to the point, who they are not.

\item Provide funding and other resources to assess the effectiveness of DEIA programs. Support collaborative initiatives to develop best practices related to DEIA. Solving the problem of underrepresentation is not easy, but an extensive amount of research exists in other fields. Funding agencies should support and fund cross-disciplinary efforts, including collaborative efforts between physicists, social scientists, social justice activists and legal experts, to study policies and practices for making changes.

\item Establish an Office of Diversity, Equity, Inclusion, and Accessibility in the agencies to closely work with different Science Offices/Divisions to set strategic priorities to advance DEIA and to develop ethical and equitable review procedures. 
\end{itemize}

\paragraph{Recommendations for Professional Societies}
\begin{itemize}

\item Encourage and aid organizations and institutions in collecting quantitative and qualitative demographic and climate data in an anonymous, ethical and informative manner. Encourage and aid organizations and institutions in using this data to plan and evaluate efforts to recruit and retain members of marginalized groups.
\item Encourage and facilitate cross-institutional data collection and evaluation.
\end{itemize}

\paragraph{Recommendations for Institutions and Collaborations}
\begin{itemize}
\item Collect quantitative and qualitative demographic and climate data, and use this data to plan and evaluate efforts to recruit and retain members of marginalized groups. Organizations and institutions that collect demographic data related to ethnicity, sexual orientation and gender identity, and disability status should collect this data in an anonymous and ethical manner, while taking care to avoid obscuring information if it is necessary to consolidate different demographic groups.
\item Make publicly available statistics related to the effectiveness of engagement programs at the high school and undergraduate level, including REUs. 
See sections: \ref{A} and \ref{D} for details.
\end{itemize}

\paragraph{Recommendations for Individuals}
\begin{itemize}
    \item Small statistics are often cited as sufficient reason to ignore problems in representation at an institutional level. But all of these small statistics together tell a consistent and statistically significant story. We invite all members of our community to reflect, learn, and take action.
\end{itemize}

\subsection{Engage Marginalized Communities}

\paragraph{Recommendations for Funding Agencies}

\begin{itemize}
\item Make publicly available statistics related to the effectiveness of engagement programs at the high school and undergraduate level. 
See sections: \ref{A} and \ref{D} for details.

\item Through a new grant, support researchers who wish the provide regular and ongoing engagement to schools and school districts in marginalized communities. See section: \ref{A} for details.
\end{itemize}

\paragraph{Recommendations for Professional Societies}
\begin{itemize}
\item As it was difficult to find an up-to-date list of successful engagement programs at the elementary through high school level, we ask professional organizations to consider promoting or facilitating an ongoing consortium of known programs. See section: \ref{A} for details.  
\end{itemize}

\paragraph{Recommendations for Institutions}
\begin{itemize}
    \item Train scientists in science communication.
    \item Build lasting relationships with marginalized communities by partnering to achieve community goals.
\end{itemize}

\paragraph{Recommendations for Scientific Collaborations}
\begin{itemize}
    \item Actively engage with Minority-Serving Institutions, and ensure the partnerships are genuine and beneficial to the MSIs.
\end{itemize}

\paragraph{Recommendations for Individual Scientists}
\begin{itemize}
\item Work to develop cultural competence in public engagement. 
\item Use principles of inclusive science communication. 
\item If you wish to make a lasting and continual impact on your local area, consider ``adopting" schools and/or school districts in marginalized communities as a resident scientist. (Relatedly, we recommend funding agencies help offset costs associated with this through the formation of a new grant.) See section: \ref{A} for details.
\end{itemize}

\subsection{Create Infrastructure to Support, Recruit, Retain, and Advance Marginalized Individuals}

\paragraph{Recommendations for Funding Agencies}
\begin{itemize}
\item Fund and support new programs aimed at marginalized communities, particularly at the primary school level. See section: \ref{A} for details.
\item The majority of community college students are from historically marginalized groups. Fund and support more programs that provide research experience to community college students who may be interested in pursuing physics degree programs at a 4-year institution. See section: \ref{B} for details.
\end{itemize}

\paragraph{Recommendations for Professional Societies}
\begin{itemize}
\item Provide resources or a networking group that focus on community college students, many of whom are from marginalized communities, who wish to pursue careers in physics. See section: \ref{B} for details.
\item Put together a list of schools that currently host or are planning to host sponsored bridge programs. See section: \ref{C} for details.
\end{itemize}

\paragraph{Recommendations for Institutions}
\begin{itemize}
\item Form partnerships with local community colleges to provide opportunities to community college students, many of whom are from marginalized communities, who wish to pursue careers in physics. See section: \ref{B} for details.
\item Make public statistics related to the effectiveness of bridge programs. See section: \ref{C} for details.
\item Encourage networking and continued mentorship for REU participants through their undergraduate to graduate career and beyond. See section: \ref{D} for details.
\end{itemize}

\paragraph{Recommendations for Collaborations}
\begin{itemize}
\item Provide accessible training, mentoring and support networks to members of marginalized groups.
\item Offer additional resources to scientists who have little local support.
\item Provide funding to relieve financial stresses and burdens placed on marginalized scientists.
\item Encourage networking and continued mentorship for high school or undergraduate student interns working within your collaboration through their graduate careers and beyond. See sections: \ref{A}  and \ref{D} for details.
\end{itemize}

\paragraph{Recommendations for Individuals}
\begin{itemize}
\item Encourage networking and continued mentorship for your student interns and researchers, particularly those you encounter through high school and undergraduate engagement programs. See sections: \ref{A} and \ref{D} for details.
\end{itemize}

\subsection{Create an Environment Conducive to Access and Success for Marginalized Individuals}

\paragraph{Recommendations for Funding Agencies}
\begin{itemize}
\item Set requirements for PIs to exclude from eligibility for grant funding those who have committed harassment or discrimination.
\item Establish a path for those funded through PI grants to report instances of harassment, discrimination, and exploitation directly to the funding agency in order to eliminate the conflict of interest created by self-policing of academic institutions.
\item Design policies that protect the flexibility and adaptability of the research environment that enables neurodiverse individuals and individuals with disabilities to make creative and innovative contributions while minimizing the burden of their symptoms.
\end{itemize}

\paragraph{Recommendations for Professional Societies}
\begin{itemize}
\item Find ways to support physicists' intersectional identities with identity-based conferences and organizations.
\item Step in to enact policies to address cross-institutional misconduct.
\item Provide support and communication platforms to help and encourage institutions and individuals with implementing DEIA policies and guidelines.
\end{itemize}

\paragraph{Recommendations for Institutions}
\begin{itemize}
\item Enact formal policies and guidelines regarding misconduct, including but not limited to harassment, discrimination, and bullying. Train students, faculty, and scholars in these policies, and provide effective structures for reporting and investigating misconduct. 

\item Advertise and encourage access to and expansion of disability accommodations. Provide additional flexibility for students/academics with disabilities to develop individualized accommodations specific to their needs.

\item Provide supports for those experiencing harassment, discrimination, and bullying that are separate from the reporting and investigation process, ensuring that this responsibility does not fall to overburdened campus counseling and mental wellness centers.

\item \sloppy{Address gaps in the enforcement of misconduct guidelines by addressing cross-institutional misconduct. Enact policies to prevent institutions from participating in the ``pass the harasser" phenomenon. Use policies at the University of California, Davis and University of Wisconsin as models.}

\item Support studies and implement guidelines regarding establishing and improving the environment to better retain students and scholars from marginalized communities.
 \item Address biases in evaluation and selection processes. Recognize the value of service work, which is disproportionately shouldered by members of marginalized communities.
\end{itemize}

\paragraph{Recommendations for Scientific Collaborations}
\begin{itemize}
  \item Adopt a code of conduct, and set guidelines for creating a welcoming environment.
  \item  Enact effective policies addressing cross-institutional misconduct or misconduct within a member institution. Scientific collaborations should provide clear guidance, keeping legal considerations in mind, about how to deal with misconduct and about who bears responsibility for investigating and disciplining member scientists reported for misconduct.
   \item Offer training to raise awareness of microaggressions that can happen in collaboration settings (telecons, messaging applications, etc.). Teach collaboration members microintervention strategies.
   \item Offer training and opportunities for education as well as open discussion of ableism in academia and of the impacts and behavioral implications of disabilities, visible and invisible.

\item Continue to support maximal flexibility and adaptability in the work environment of collaborators in order to enable neurodiverse individuals to thrive in the research environment. 

\item Continue to encourage knowledge sharing, open-access to information, and a spirit of true collaboration in order to reduce entry-level barriers which disproportionally hamper marginalized individuals to make scientific contributions.
\end{itemize}

\paragraph{Recommendations for Individuals}
\begin{itemize}
 \item All members of our community, but particularly those with more privilege, should participate in learning and widening DEIA efforts to make change. Study the existing literature to gain a better understanding before asking marginalized people to do extra work. It is unsustainable for marginalized groups to carry out this important work alone.
\end{itemize}


\subsection{Outlook for the Next Snowmass}

By the next Snowmass process, we hope to see improvement to the overall climate of particle physics, as described in the Early Career Survey white paper \citep{2022arXiv220307328A}. We hope to see particle physics become more diverse, equitable, inclusive and accessible. We also hope to see statistics on how many of the recommendations outlined in this and other white papers have been implemented. Finally, as has been pointed out in this and other Snowmass white papers \citep{hansensmith_climatecp}, and has been demonstrated in the process of writing those white papers, DEIA-focused tasks often disproportionately fall upon people from marginalized groups. In the next Snowmass process, we hope that the demographics of participants in DEIA-related activities will better reflect the demographic distribution of all Snowmass participants. In addition, we hope that our community will soon establish an entity, possibly within the Division of Particles and Fields, to monitor the progress toward diversity, equity, inclusion, and accessibility. 
